\documentstyle[12pt,nature,psfig]{article}
\begin{document}
\newcommand{\be}{\begin{eqnarray}}
\newcommand{\ee}{\end{eqnarray}}
\newcommand{\etal}{{\it{et al.}}}
\newcommand{\smass}{M_{\odot}}
\newcommand{\br}{{\bf r}}
\newcommand{\bV}{{{\bf v}}}
\renewcommand{\thefootnote}{\alph{footnote}}

\newcommand{\kms}{\mbox{${\rm km~s}^{-1}$}}
\newcommand{\msun}{\mbox{${\rm M}_\odot$}}
\def\apgt{\ {\raise-.5ex\hbox{$\buildrel>\over\sim$}}\ }
\def\aplt{\ {\raise-.5ex\hbox{$\buildrel<\over\sim$}}\ }

%\baselineskip 24pt

%% You can insert a short comment on the title page using the command below.
%\slugcomment{Not to appear in Nonlearned J., 45.}

\def\steve#1{{\bf[#1 -- Steve]}}
\def\Steve#1{{\bf[#1 -- Steve]}}
\def\simon#1{{\bf[#1 -- Simon]}}
\def\Simon#1{{\bf[#1 -- Simon]}}
\def\holger#1{{\bf[#1 -- Holger]}}
\def\Holger#1{{\bf[#1 -- Holger]}}
\def\Jun#1{{\bf[#1 -- Jun]}}

%\shorttitle{}
%\shortauthors{Portegies Zwart {\it et al.}}
%
% Title
%
\title{
The formation of massive black \\
holes through collision runaway \\
in dense young star clusters
}

%
% Authors
%

\author{
Simon F. Portegies Zwart\\
Astronomical Institute `Anton Pannekoek', \\
		University of Amsterdam, Kruislaan 403 \\
{\small and}\\
Institute for Computer Science,\\
		 University of Amsterdam, Kruislaan 403 \\
{\small{spz@science.uva.nl}}\\
~\\
Holger Baumgardt\\
 Institute of Advanced Physical and Chemical Research  RIKEN,\\
   2-1 Hirosawa, Wako-shi, Saitama 351-019, Japan\\
~\\
Piet Hut\\
Institute for Advanced Study,
Princeton, NJ 08540, USA\\
~\\
Junichiro Makino\\
Department of Astronomy, University of Tokyo, Tokyo 113, Japan\\
~\\
Stephen L.\ W.\ McMillan\\
Department of Physics,
		 Drexel University,\\
                 Philadelphia, PA 19104, USA\\
}
%
% Abstract
%

\def\jun#1{{\bf[#1---Jun]}}
\def\piet#1{{\bf[#1---Piet]}}

\maketitle

\begin{abstract}
\noindent{\bf A luminous X-ray source is associated with a cluster
(MGG-11) of young stars $\sim 200$\,pc from the center of the
starburst galaxy M82 \cite{Matsumoto1999,Kaaret2001}. The properties
of the X-ray source are best explained by a black hole with a mass of
at least 350\,\msun\, \cite{Matsumoto2001,Strohmayer2003}, which is
intermediate between stellar-mass and supermassive black holes.  A
nearby but somewhat more massive star cluster (MGG-9) shows no
evidence of such an intermediate mass black hole
\cite{Matsumoto1999,Matsumoto2001}, raising the issue of just what
physical characteristics of the clusters can account for this
difference. Here we report numerical simulations of the evolution and
the motions of stars within the clusters, where stars are allowed to
mergers with each other. We find that for MGG-11 dynamical friction
leads to the massive stars sinking rapidly to the center of the
cluster to participate in a runaway collision, thereby producing a
star of 800--3000\,\msun, which ultimately collapses to an black hole
of intermediate mass. No such runaway occurs in the cluster MGG-9
because the larger cluster radius leads to a mass-segregation
timescale a factor of five longer than for MGG-11.  }
\end{abstract}

\begin{figure}[htbp!]
\psfig{figure=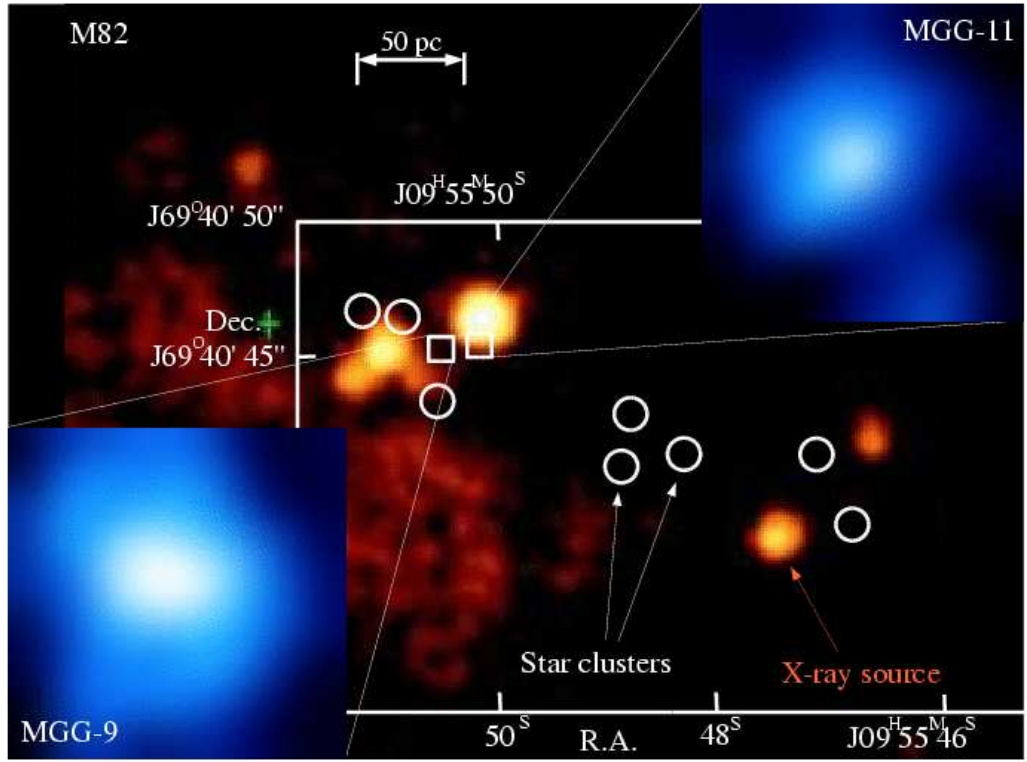,width=\columnwidth} 
%Figure {\bf Figure1.ps} should come here
\caption[]{ 
{\em Chandra} X-ray image (in R.A. and Dec.) of the
relevant part of the starburst galaxy M82 with the observed star
clusters indicated.  The color image is from the 28 October 1999 X-ray
observation by Matsumoto et al.\, (2001),\cite{Matsumoto2001} with
about arcsecond positional accuracy.  The brightest X-ray source (M82
X-1) is near the center of the image.  The star clusters, from table 3
of McCrady et al. (2003),\cite{McCrady2003} are indicated by circles.
The positions of the two star clusters MGG-9 and MGG-11 are indicated
with squares. The magnified infrared images of these star clusters
from the McCrady et al observations are presented in the upper right
(MGG-11) and lower left (MGG-9) corners.  A recently discovered
$54.4\pm0.9$\,mHz quasi-periodic oscillator is not shown because of
its low (7 arcsecond) positional accuracy\cite{Strohmayer2003}, but
its position is consistent with the X-ray source in MGG-11. A
millimeter source\cite{Matsushita2000} roughly centered around the two
clusters is not shown either, because it is a large shell-like
structure with a diameter of $\sim 14'' \times 9''$.
}
\label{fig:obs}
\end{figure}

Using the Keck NIRSPEC spectrometer, McCrady et al.\cite{McCrady2003}
have made accurate measurements of the bulk parameters of MGG-11 and
MGG-9, two of the brightest star clusters in the central region of
M82.  Figure\,\ref{fig:obs} shows X-ray\cite{Matsumoto2001} and
near-infrared\cite{McCrady2003} images of the area of interest around
the two clusters.  The relative positional accuracies of both sets of
observations are better than 1 arcsecond.  However, the absolute
pointing accuracy is much poorer, for both telescopes.  Although
apparently off-center, the positions of the star cluster MGG-11 and
the bright X-ray source are in fact consistent with one another
(D. Pooley, private communication).

Both MGG-9 and MGG-11 have quite similar ages, in the range
7--12\,Myr \cite{McCrady2003}. The line-of-sight velocity dispersion of
MGG-11 ($\sigma_r = 11.4\pm0.8\,\kms$) is somewhat smaller than that
of MGG-9 ($\sigma_r = 15.9\pm0.8\,\kms$).  Combining these numbers
with the projected half-light radii, $r = 1.2$\,pc for MGG-11
and $r = 2.6$\,pc for MGG-9, McCrady et al.~estimate a total
cluster mass of $\sim (3.5\pm 0.7) \times 10^5 \msun$ for MGG-11, and
about four times higher for MGG-9.

In young dense clusters like MGG-11, supermassive stars may form
through repeated collisions \cite{Portegies Zwart1999, Portegies
Zwart2002, Gurkan et al 2003}. The collision rate will be greatly
enhanced if massive stars have time to reach the core before exploding
as supernovae \cite{Ebisuzaki2001}. Dynamical friction implies a
characteristic time scale $t_{\rm df}$ for a massive star in a roughly
circular orbit to sink from the half-mass radius $R$ to the cluster
center\cite{Spitzer1971}:
\begin{equation}
      t_{\rm df} \simeq {\langle m \rangle \over 100\msun} 
                        {0.138 N \over \ln(0.11 M/100\msun) } 
                        \left( R^3 \over GM \right)^{1/2}.
\label{Eq:tdf}
\end{equation}
Here $\langle m \rangle$ and $M$ are the mean stellar mass and the
total mass of the cluster, respectively, $N$ is the number of stars,
and $G$ is the gravitational constant. 
For definiteness, we have evaluated $t_{\rm df}$ for
a 100\,{\msun} star.  Less massive stars undergo weaker dynamical
friction, and thus must start at smaller radii in order to reach
the cluster center on a similar time scale.

For MGG-11, we find $t_{\rm
df}\sim 3$\,Myr, which is comparable to the main-sequence lifetimes of
the most massive stars.  On the other hand, for MGG-9, $t_{\rm df}\sim
15 $\,Myr.  Thus, massive stars in MGG-11 can easily reach the center
of the cluster before exploding as supernovae, whereas those in MGG-9 do not.
Given the high central density of MGG-11, its massive stars, once
accumulated in the cluster center, cannot avoid a runaway
collision \cite{Portegies Zwart1999}.

We have tested this scenario by carrying out star-by-star simulations
of MGG-11 and MGG-9 using two independently developed $N$-body codes,
{\tt Starlab}\cite{PortegiesZwart2001,starlab} and {\tt
NBODY4} \cite{Aarseth1999, Baumgardt2003}. A more extensive discussion
of those simulations is presented as {\em Supplementary
Information}. Briefly, we choose the initial conditions of our model
clusters so that today, at an age of 7--12\,Myr, they have mass
functions, luminosities, half-mass radii and velocity dispersions in
agreement with the McCrady et al.~observations.  Since we do not know
either the initial or the current central densities of either cluster,
the concentration $c$ is treated as a free parameter controlling the
initial central density of our model clusters.  [The concentration
parameter is defined as the ratio of the core radius to the tidal
radius: $c \equiv \log(R_{\rm t}/R_{\rm c})$.]

\begin{figure}[htbp!]
\psfig{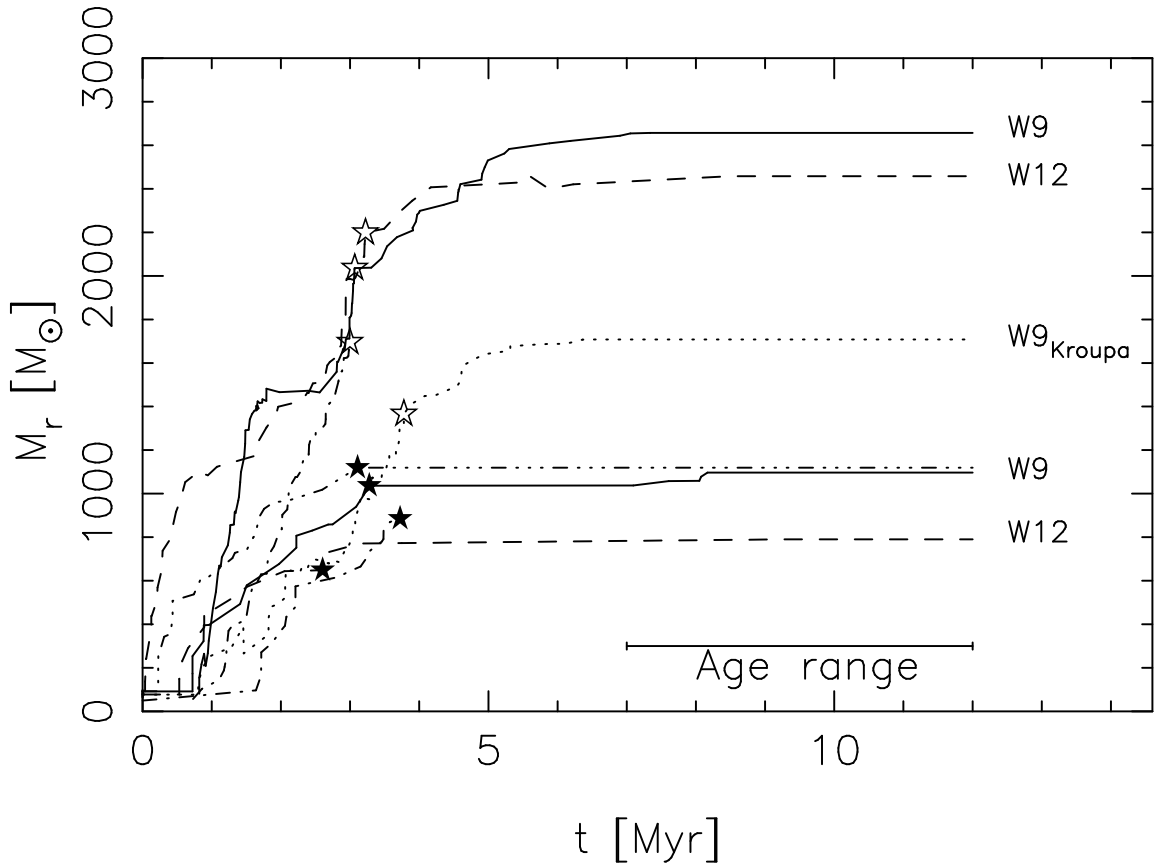} 
%Figure {\bf Figure2.ps} should come here
\caption[]{ 
The growth in mass of the collision runaway star with
time.  The choice of initial concentration is labeled by the central
potential, where W12(9) implies $W_0 = 12(9)$.  For both choices, the
top curves give the {\tt NBODY4} results, and the bottom curves the
{\tt Starlab} results.  The runaway masses in {\tt NBODY4} are larger,
since that code adopts larger stellar radii, as discussed in the {\em
Supplementary material}.
The star symbols indicate the moment when the runaway experiences a
supernova, typically around 3\,Myr.  The open and filled stars indicate
simulations performed with {\tt NBODY4} and {\tt Starlab}, respectively.
The solid and dashed curves show $M_r$ for a
Salpeter\cite{Salpeter1955} IMF with a lower limit of 1\,{\msun} and
$c \simeq 2.1$ ($W_0 = 9$) and $c \simeq 2.7$ ($W_0=12$).  The
dash-dotted curves are for two models with $W_0=9$ with an upper limit
to the IMF of 50\,{\msun}, instead of the standard 100\,{\msun} used in
the other calculations; we terminated these runs at the moment the
runaway star experiences a supernova.  The dash-3-dotted curve shows
the result for $W_0=12$ with a Salpeter IMF and with 10\% primordial
binaries. Finally, the dotted curve shows results for $W_0=9$ and a
Kroupa\cite{Kroupa2001} IMF with a minimum mass of 0.1\,\msun, in a
simulation with 585,000 stars.
The observed age range of MGG-11 and MGG-9 is indicated by the
horizontal bar near the bottom of the figure.
}
\label{fig:Mbh}
\end{figure}

We find that, for $c > 2$ (which for ``King''\cite{King1966} models is
equivalent to a dimensionless central potential $W_0 \apgt 9$) our
MGG-11 models do indeed show runaway growth via repeated collisions.
Figure \ref{fig:Mbh} presents, for several such simulations, the
growth in mass of the star that would ultimately become the most
massive star in the cluster.  We will refer to this star simply as
``the runaway star''. Based on detailed supernova calculations, we
assume that stars having masses $> 260$\,{\msun} collapse to black
holes without significant mass loss in supernova
explosions\cite{Heger2003}.  Our stellar evolution models for stars
with masses between 50 and 1000\,{\msun} are based on detailed
calculations for such high-mass stars \cite{Stothers1997, Ishii1999}.

The quantitative differences evident in Figure \ref{fig:Mbh} between
the simulations performed with {\tt Starlab} and those using {\tt
NBODY4} are due mainly to the different
radii assumed for the runaway star in those two packages.  Stars of
masses $\apgt 100$\,{\msun} in {\tt NBODY4} are larger (in a
time-averaged sense) by about a factor of 3
compared to those in {\tt Starlab}. (More details are provided in {\em
Supplementary Information}.)  Because gravitational focusing
dominates the collision cross section, this difference propagates
linearly in the collision rate, explaining the factor $\sim 3$
difference in the final mass of the runaway star.  Apart from
this effect,  we find
that our qualitative results are quite insensitive to the details of
the adopted evolution prescription.

For MGG-11, the runaway star typically experienced a total of
$\sim10$--100 collisions.  Most of them occurred during the first 3
Myr, that is, before the star became a black hole.  The collision
counterparts are usually 30--50 {\msun} main-sequence stars. By the
time the runaway star collapsed to a black hole it had reached a mass
of 800--3000\,\msun. Later the black hole may capture a companion star
to become an ultraluminous X-ray source \cite{Hopman2003}.

No episode of runaway growth was seen in our MGG-11 models with $c <
2$, nor in any of the MGG-9 simulations, regardless of initial
concentration. Thus we see very clearly that differences in bulk
parameters, specifically the dynamical friction time scale for the
most massive stars, can readily explain why MGG-11 might host an IMBH
while MGG-9 does not.  Simulations of MGG-11 with $c > 2$ ($W_0 \apgt
9$) reached core collapse\cite{HH2003} before 1Myr.  Models with $c <
2$ ($W_0 \aplt 8$) also showed some increase in central density, but
the maximum density was much lower (and occurred later) because the
collapse was stopped by supernova mass loss after $\sim 3$ Myr. Thus,
a collision runaway could not occur in these clusters.

\begin{figure}[htbp!]
\psfig{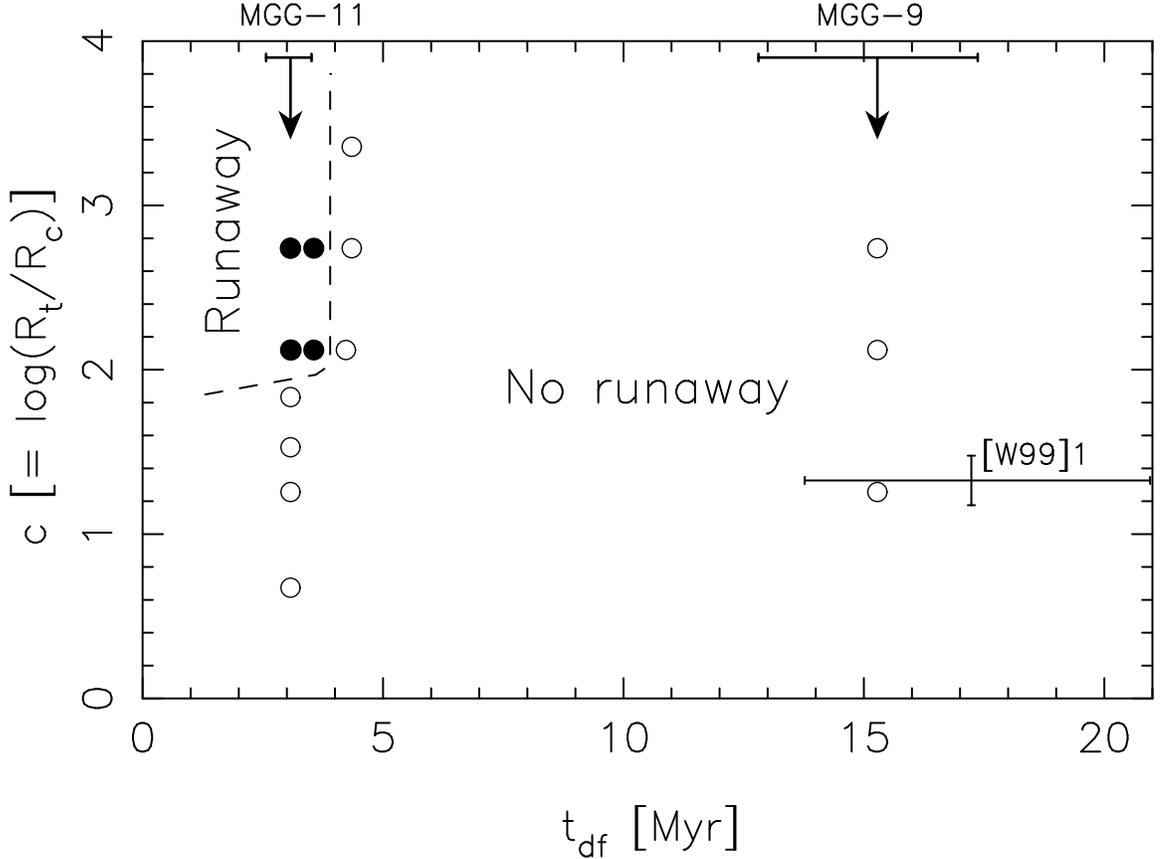}
%Figure {\bf Figure3.ps} should come here
\caption[]{ 
The area of parameter space for which runaway collision
can occur, and where the process is prevented.  Conditions for runaway
merging identified through our simulations in the \{$t_{\rm df}$,
$c$\} plane, where $t_{\rm df}$ is the dynamical friction time scale
for 100\,{\msun} stars, and $c$ the cluster concentration
parameter. The horizontal error bars for the initial conditions for
MGG-9 and MGG-11 reflect errors in the observed cluster mass and
projected half light radius \cite{McCrady2003}. The best fit for the
concentration parameter of the Antennae star cluster [W99]1 is
indicated by a vertical bar near $t_{\rm df} \simeq 17$\,Myr, the
horizontal error bar reflect the uncertainty in the measured cluster
mass and radius \cite{Mengel2002}. Solid circles indicate simulations
resulting in runaway merging, while open circles correspond to
simulations in which no runaway merging occurred.  The evolution of
the mass of the runaway for the two leftmost filled circles are
presented in Figure \ref{fig:Mbh}, for a variety of initial
conditions.  Star clusters in the upper left corner enclosed by the
dashed line are expected to host an IMBH formed by the runaway growth
of a single star.  In the other parts of the diagram (lower
concentration and/or larger dynamical friction time) an IMBH cannot
form by this process.  }
\label{fig:tdf_c}
\end{figure}

Figure\,\ref{fig:tdf_c} summarizes the results of our simulations,
illustrating how both high initial concentration and short dynamical
friction time scales are needed to lead to a collision runaway.  In
addition to the parameters for MGG-9 and MGG-11, we also indicate on
the figure the location of the young star cluster
[W99]1\cite{Whitmore1999}, one of several star clusters in the
Antennae system for which accurate structure parameters have been
determined \cite{Mengel2002}. Its parameters are quite similar to those
of MGG-9.  The cluster [W99]1 has a bright $L_{0.2-10{\rm keV}} \simeq
10^{38.7}$ erg/s X-ray point source as a counterpart,\cite{Zezas2002}
the luminosity of which is consistent with a ``normal'' high-mass
X-ray binary containing a magnetized neutron star or a stellar-mass
black hole \cite{Zezas2002}. Star clusters such as MGG-11, MGG-9 and
[W99]1 are richly populated with black holes and neutron stars---based
on our adopted mass function, we expect these clusters to contain some
1400 stellar-mass black holes and up to about 1000 neutron stars. The
formation of an ordinary high-mass X-ray binary is therefore not
surprising.

The relevant parameters of five other young clusters in the
Antennae\cite{Mengel2002} all fall far to the right of
Figure \ref{fig:tdf_c} and are therefore not expected to contain IMBHs.
This is consistent with the absence of ultraluminous X-ray sources in
these clusters.  The Milky Way galaxy contains (at least) four young
dense star
clusters, of which the Arches, Quintuplet, and NGC 3603 have the right
conditions for multiple stellar collisions to occur.  However, the
relatively small number of stars in these systems may prevent the
growth of an object massive enough to collapse to an IMBH. In any
case, these clusters are currently too young to have experienced any
supernovae. The slightly older star cluster Westerlund
1\cite{Piatti1998} is sufficiently massive, and fulfills the criteria
for producing an IMBH via the process described above.  So far, no
bright X-ray source has been found in this cluster.

Based on theoretical considerations, supported by extensive numerical
simulations, we have shown that an IMBH of mass
$\sim$800--3000\,{\msun} is expected to form through runaway
collisions in the star cluster MGG-11, but not in MGG-9.  The
requirement for the formation of such an IMBH in MGG-11 is that the
cluster was born with $c \apgt 2$ and $t_{\rm df} \aplt 4$Myr.  While
high by the standards of typical open and globular clusters observed
in our Galaxy, such a density is not uncommon among young ($\aplt
3$\,Myr old) star clusters.  Examples include NGC\,3603 ($c \ge
2.08$)\cite{Moffat et al 1994} in the Milky Way Galaxy, and R\,136 in
the 30 Doradus region of the LMC \cite{Campbell et al 1992}. We propose
that the IMBH thus formed is the origin of the $L_{0.2-10keV} \simeq
10^{41}$ erg/s ultraluminous X-ray point source seen at the position
of MGG-11.
 
\newpage

We are grateful to Nate McCrady and Dave Pooley for discussions on
MGG-11, to Takeshi Go Tsuru for an accurate position of M82 X-1 and to
Edward van den Heuvel for critically reading the manusctipt.  This
work was supported by NASA ATP, the Royal Netherlands Academy of
Sciences (KNAW), the Dutch organization of Science (NWO), and by the
Netherlands Research School for Astronomy (NOVA).

\bigskip

Correspondence and requests for materials should be addressed to Simon
Portegies Zwart (email: spz@science.uva.nl)

\vfill
\newpage

\vfill
\newpage

\section*{Supplementary material}

{\large \bf This part of the paper will be published online and it has
a separate bibliography }

In this section we discuss the selection of initial conditions,
numerical methods, and the results of our simulations of the star
clusters MGG-9 and MGG-11.

The first step in modeling this problem is to reconstruct the
conditions under which MGG-11 was born 7--12\,Myr ago.  McCrady et
al\cite{McCrady2003} determine the cluster age and mass-to-light
ratio, and from these conclude that the mass function is consistent
with a power-law having the Salpeter\cite{Salpeter1955} slope and a
lower mass limit of $\sim1\,\msun$ \cite{McCrady2003}. At the current
age of the cluster, all stars more massive than
$\sim$17--25\,\msun\cite{Eggleton1989} have experienced supernovae.
We adopt this mass function in most of our simulations, with an
upper limit of 100\,{\msun}.  This choice of mass limit is motivated
by the fact that there are several clusters in the Milky Way Galaxy
with characteristics similar to the two clusters studied here (MGG-11
and MGG-9) that are young enough for the highest-mass stars still to be
present. These clusters are the Arches\cite{Figer},
Quintuplet\cite{Figer1995}, NGC3603\cite{Nurnberger} and the central
cluster R136\cite{MH1998} in the 30 Doradus region in the large
Magellanic cloud (LMC).  The publications cited here support the existence
of stars of $\apgt100$\,{\msun} in these clusters.  In addition, an
upper mass limit of $\sim 100\,\msun$ in the Galactic disc is probably
quite realistic \cite{Kroupa and Weitner}.

The 1\,{\msun} lower mass cut-off for the initial mass function (IMF)
is somewhat controversial. However, the mass functions in several of
the young star clusters in the Antennae galaxies\cite{Mengel2002}, the
Arches cluster\cite{Figer}, Quintuplet cluster\cite{Figer1999},
NGC1705-1\cite{Smith and gallagher} and the star cluster MGG-F in
M82\cite{Smith and gallagher} are consistent with a similar lower mass
cut-off. For our study we adopt the observed mass function as a first
choice, but in later runs we vary the mass function to test the effect
of changing it.

At zero age the mean mass of our adopted IMF is $\langle m \rangle
\simeq 3.1\,M_\odot$.  At the observed age of 7--12\,Myr the mean
stellar mass has dropped to $\langle m \rangle \simeq 2.5-2.7\,
\msun$\, \cite{Eggleton1989}. What actually happens to stars of
$\sim$17--25\,{\msun} which experience a supernova before 7--12\,Myr
is not crucial for this estimate, as there are only 1500--2500 such
stars.  If we naively assume that all the mass in these stars is
simply lost from the cluster we obtain mean stellar mass of
2.5--2.6\,\msun.  If we assume that stars more massive than
25\,{\msun} turn into 10\,{\msun} black holes and stars of lower mass
into Chandrasekhar [1.4\,\msun] mass neutron stars, the mean stellar
mass drops to $\langle m \rangle \simeq 2.6-2.7 \msun$.  With today's
measured mass of $M = 3.5\times 10^5$\,\msun\cite{McCrady2003} the
cluster then contains $N = 130,000$--140,000 stars, implying an
initial total mass of $M_0 = 4.1 $--$4.5 \times 10^5$\,\msun.

In order to model the cluster we must know its initial half-mass
radius $R$.  We obtain this parameter by de-projecting the observed
half-light radius and correcting for the expansion of the cluster
during its lifetime.  The two corrections are comparable, but with
opposite sign: the half-mass radius is $\sim$25 percent larger than
the observed half-light radius, and mass loss due to stellar evolution
drives adiabatic expansion of the cluster radius by $\sim$20
percent. The initial half-mass radius of MGG-11 is then $R \simeq
1.2$\,pc. Here we assume that mass follows light after 12 Myr,
i.e.~that the light profile is not significantly affected by the
segregation of the most massive stars. This turns out to be a
reasonable assumption.  Note that the mass-to-light ratio for MGG-11
reported by McCrady et al. is $L/M = 3.5^{+2.1}_{-1.5}$, and that
adding a 3000\,{\msun} non-luminous object to the cluster increases
this value only by about 1 percent.  A similar analysis for MGG-9
results in a total mass at birth of $M_0 \simeq 1.8 \times
10^6$\,{\msun} and an initial half-mass radius of $R \simeq 2.6$\,pc.

We ignore the effect of the tidal field of M82 in our simulations.
The runaway growth studied for these clusters takes place within the
dynamical friction time scale of the most massive stars and the
process discussed here takes place deep within the cluster core. Even
though the cluster may lose a few stars to its host Galaxy, this does
not affect the processes driving the collision runaway.  Note that the
two clusters MGG-9 and MGG-11 are close together on the sky, and could
in principle be bound in a binary cluster.  At the
3.9\,Mpc\cite{Sakai1999} distance of M82, the projected distance
between MGG-9 and MGG-11 is about 30\,pc. We note that in the LMC
there are a few hundred binary clusters for which the projected
separation is smaller than 20 pc \cite{Bica, Dieball}.

Our simulations were performed using two independently developed
$N$-body packages, {\tt Starlab}\cite{PortegiesZwart2001,starlab} and
{\tt NBODY4} \cite{Aarseth1999, Baumgardt2003}. Both codes achieve
their highest speed when running on the GRAPE-6 special-purpose
hardware \cite{Makino1997, Makino2003}.  A comparison between {\tt
Starlab} and {\tt NBODY4} is presented by Heggie
(2001)\cite{KyotoII}. % and Aarseth (2003) \cite{Aarseth2003}.

Both codes incorporate parametrized stellar evolution and allow for
the possibility of direct physical collisions between stars, thus
including the two key physical elements in our runaway scenario.  The
criteria for stellar collisions in both codes are quite similar; a
collision occurs if the distance between two stars becomes smaller
than the sum of the stellar radii. During a collision mass and
momentum are conserved.  These are reasonable assumptions since the
relative velocity of any two colliding stars is typically much smaller
than the escape speed from either stellar surface \cite{Lombardi}.  As
the runaway star grows in mass its angular momentum increases due to
off-axis collisions.  Our stellar evolution prescriptions did not
incorporate special treatments to account for this effect. Rapidly
rotating stars tend to develop significant horizontal turbulence
which may dramatically affect the further evolution of the star
\cite{2001ApJ...548..323S,2003A&A...399..263M}.  Calculations for a
rapidly rotating star of 120\,\msun\, results in a $\sim 15$ percent
extension of the lifetime compared with a non-rotating star
\cite{2003A&A...404..975M}.  This effect is stronger for lower mass
stars, but has not yet been explored for more massive stars.

The collision radius of a star is taken to be identical to the stellar
size, except for black holes for which we use the tidal radius, which
is considerably larger than the Schwarzschild radius. The zero-age
radii of various stellar evolution models are presented in Figure
\ref{fig:MR}. The radii in {\tt Starlab} are systematically smaller
than those in {\tt NBODY4}.  The collision cross sections in the
appropriate range of initial conditions are dominated by gravitational
focusing, and therefore the stellar size propagates linearly in the
collision rate, and consequently also in the final mass of the runaway
star.

We simulated MGG-11 and MGG-9 using initial mass functions and half
mass radii based on the observations of McCrady et
al.\cite{McCrady2003} Most of these calculations were performed with
131072 (128k) stars. The initial density distribution is poorly
constrained by the observations and we therefore performed a series of
simulations with various choices for the initial density profile,
ranging from the very shallow $c \simeq 0.67$ (equivalent to $W_0=3$)
to highly concentrated $c \simeq 3.4$ (equivalent to $W_0=15$). To
further test our theoretical understanding that a collision runaway is
prevented by the first supernova explosions for $t_{\rm df} \apgt
4$\,Myr we systematically increased the initial $t_{\rm df}$ for two
selected values of $c$ in our MGG-11 simulations (by varying the
initial half-mass radius $R$), until the collision runaway was
prevented. The results are presented in Figure 3 of the main
text. Note that we plot the horizontal error bars for both clusters
near the top of the figure for practical reasons; these should not be
interpreted as upper limits on the concentration parameter.

\begin{figure}
\psfig{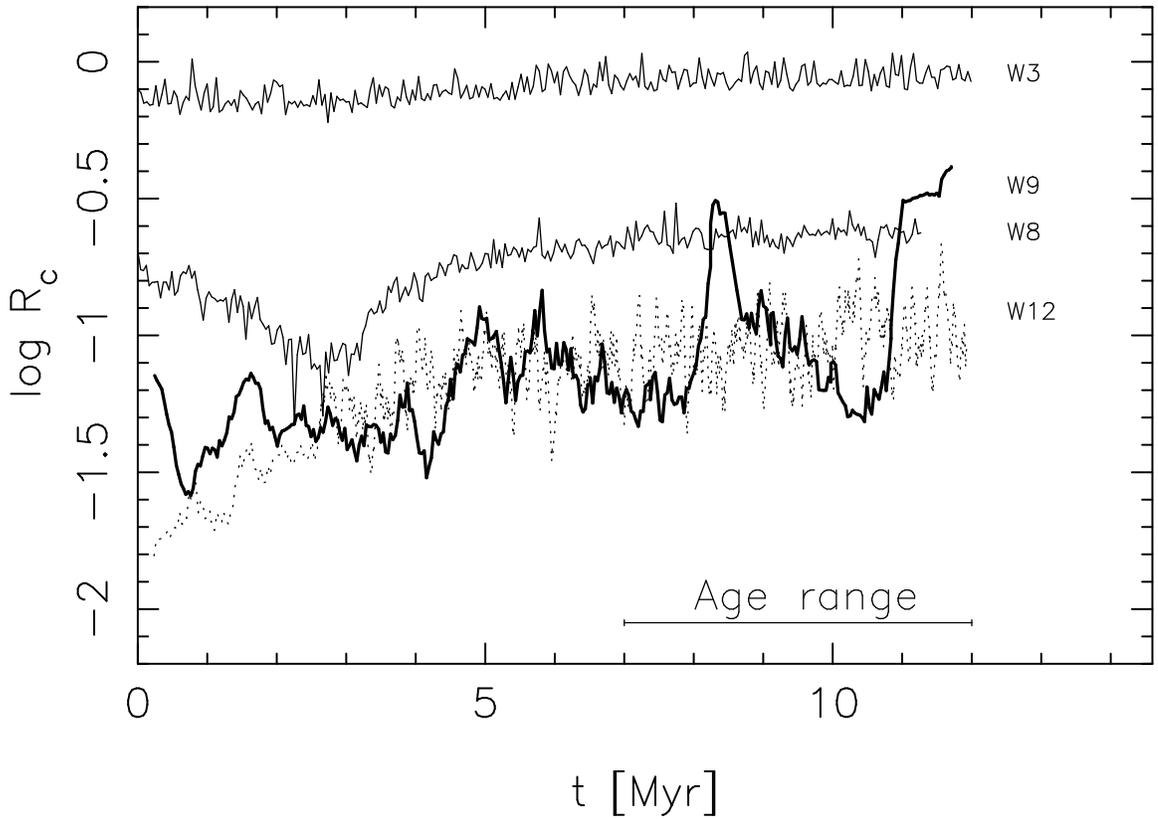}  
%Figure {\bf Figure1_SI.ps} should come here
\caption[]{
Evolution of the core radius for four simulations of the
star cluster MGG-11.  These calculations are performed with {\tt
Starlab} with $c \simeq 0.67$ ($W_0=3$), $c \simeq 1.8$ ($W_0=8$), $c
\simeq 2.1$ ($W_0=9$, bold curve) and $c \simeq 2.7$ ($W_0=12$, dotted
curve) indicated along the right edge of the figure as $W3$, $W8$,
$W9$, and $W12$ respectively.  The W9 curve is plotted with a heavy
line to distinguish it from the curves for W8 and W12. The age range
of the observed clusters is indicated near the bottom of the figure.
The large variation in the core radius for the $W_0=9$ and $W_0=12$
models is caused in part by the presence of the runaway star.  }
\label{fig:Rcore}
\end{figure}

\begin{figure}
\psfig{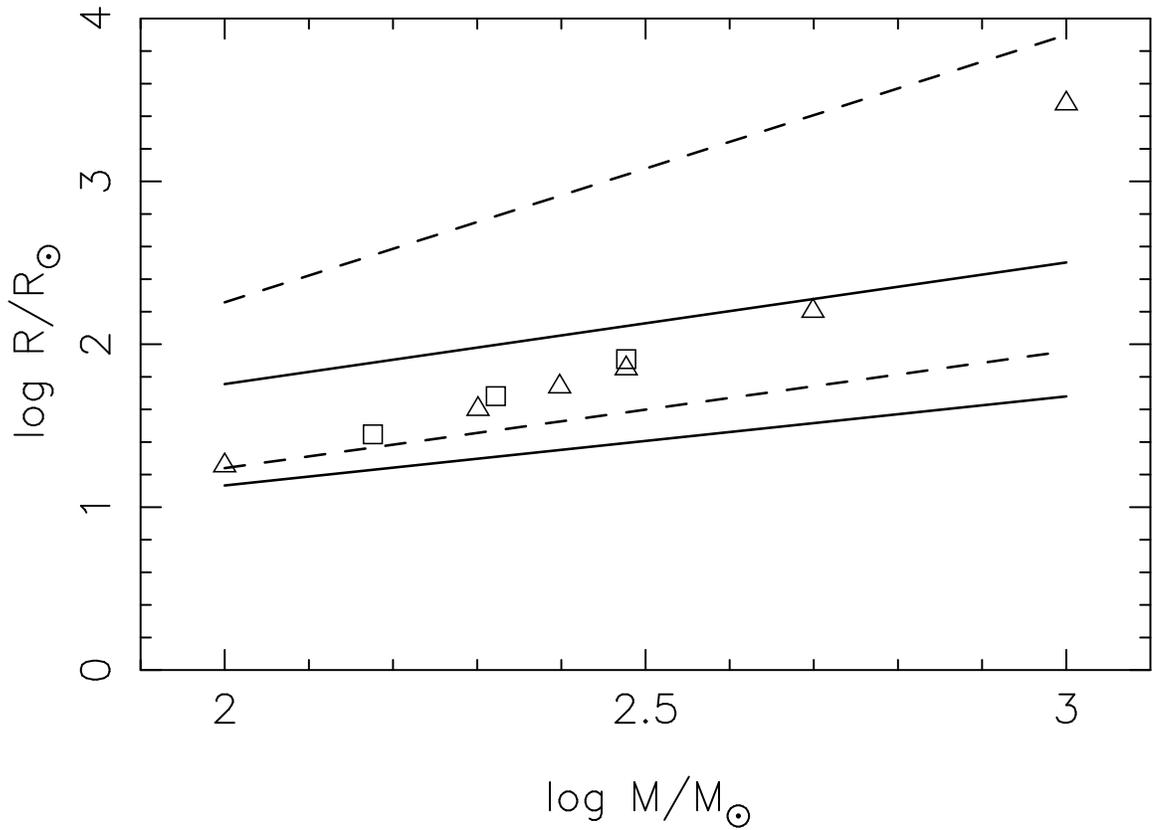} 
%Figure {\bf Figure2_SI.ps} should come here
\caption[]{
Stellar radius as a function of zero-age mass for various
models of high-mass stars.  The lower and upper solid lines indicate
the radii of zero-age and terminal age main-sequence stars from the
stellar evolution model used in {\tt Starlab}, which are based on the
prescription of Eggleton et al.\, (1989)\cite{Eggleton1989}.  The
dashed lines are taken from the stellar evolution prescription used in
{\tt NBODY4} which are based on the evolutionary calculation by Hurley
et al.\, (2000)\cite{Hurley2000}. The triangles are from the
calculations of Ishii et al.\,(1999)\cite{Ishii1999}, the squares are
taken from Stothers (1997) \cite{Stothers1997}. }
\label{fig:MR}
\end{figure}

In order to illustrate the early core collapse in the more
concentrated models, and the absence thereof in the shallower models,
Figure \ref{fig:Rcore} plots the evolution of the core radius for a
selected number of simulations for MGG-11.  We discuss each curve in
turn, starting at the top.  The core radius of the shallow model, $c
\simeq 0.67$ ($W_0=3$), hardly changes with time. The intermediate
model $c \simeq 1.8$ ($W_0=8$) almost experiences core collapse near
$t=3$\,Myr but as stellar mass loss starts to drive the expansion of
the core it never really experiences collapse.  Inspection of Figure 3
of the main text shows that neither of these clusters experienced a
collision runaway.  Core collapse occurs in the $c \simeq 2.1$
($W_0=9$) model near $t=0.8$\,Myr, which also corresponds to the time
of the first collision. The $c \simeq 2.7$ ($W_0=12$) simulation is so
concentrated that it starts virtually in core collapse, and the entire
cluster evolution is dominated by a post-collapse phase. Inspection of
Figure 2 in the main text reveals that in these calculations the
runaway also tends to start earlier.  The larger fluctuations in the
core radius for models $W_0=9$ and $W_0=12$ compared to the less
concentrated models reflect the orbit of the runaway star around the
density center of the star cluster.  The runaway star is always found
in the core, but occasionally at a substantial distance from the
center.

We performed four simulations for MGG-11 adopting $W_0=9$ with
different initial mass functions in order to study the effect of the
mass function on the calculations.  Reducing the upper limit on the
IMF from 100\,{\msun} to 50\,{\msun} reduces the mass of the runaway
star at the moment of the supernova explosion by about 20
percent. Simulations with the Kroupa (2001)\cite{Kroupa2001} IMF and a
lower mass limit of 0.1\,\msun, as well as one run with a Salpeter IMF
and a lower mass limit of 0.2\,{\msun}, resulted in a $\sim 30$\%
lower runaway mass.  These latter simulations were run with 585,000
stars. The lower mass of the runaway star in the simulations with the
reduced upper mass limit is a direct consequence of the smaller number
of massive stars in the simulation, causing the average mass of the
collision counterparts to be smaller and the collision rate to be
lower. We also performed a calculation using the initial conditions
for MGG-11 but with the zero-age stellar radii published by Ishii et
al.\,(1999)\cite{Ishii1999} for the collision runaway. This
calculation resulted in the a similar black hole mass as in the {\tt
NBODY4} calculations with otherwise the same initial conditions.

In addition, we performed several simulations for MGG-9, which is well
outside the regime where a collision runaway can occur (see Figure 3
of the main text).  For MGG-9 the IMF is consistent with a Salpeter
IMF with a lower mass limit of 0.2\,{\msun}, and also with the Kroupa
IMF.  However, due to computational limitations we could only perform
simulations with the 1\,{\msun} cut-off IMF for MGG-9. Simulations
with the Kroupa IMF would require 3.6 million stars, which is beyond
our current capabilities. In the case of the Salpeter IMF with a lower
limit of 1\,{\msun}, and with other parameters chosen consistent with
the observations of McCrady et al for MGG-9, we did not observe
runaway merging. 

Finally, we performed two calculations for MGG-11 using $c \simeq 2.7$
($W_0=12$), and including 10 percent primordial binaries.  The current
binary fractions for MGG-11 is not determined by the observations, but
based on studies of other clusters, 10 percent seems a reasonable
choice \cite{Albrow, Rubenstein, Yan}. We select initial orbital
separations ranging from Roche-lobe contact (at a binding energy of
about 300\,kT) to that corresponding to a binding energy of $\sim
3$\,kT, which is near the dividing line between hard and soft
binaries \cite{Heggie1975}. (Note that the total kinetic energy of the
stellar system is $\frac32NkT$.)  The other initial binary parameters
were selected as in Portegies Zwart et
al. (2001) \cite{PortegiesZwart2001}. Though modest, this fraction
should be sufficient to completely dominate the dynamical evolution of
the cluster core \cite{MHM1991}.

The presence of primordial binaries in our models causes an increase
in the overall merger rate by about a factor of four.  Most of these
mergers do not involve the runaway star but occur as a result of
unstable mass transfer in a primordial binary.  The collision rate
with the runaway star increases by about 30 percent, and as a consequence
the mass of the runaway star at the time of supernova is 22--36 percent
larger than without primordial binaries.  We have neglected the
possibility of primordial mass segregation, which would further favor
the runaway process, as massive stars would be born preferentially in
the cluster core.

\vfill
\newpage 

\end{document}